\documentclass[pdflatex,sn-mathphys-ay]{sn-jnl}

\usepackage{graphicx}%
\usepackage{multirow}%
\usepackage{amsmath,amssymb,amsfonts}%
\usepackage{amsthm}%
\usepackage{mathrsfs}%
\usepackage[title]{appendix}%
\usepackage{xcolor}%
\usepackage{textcomp}%
\usepackage{manyfoot}%
\usepackage{booktabs}%
\usepackage{listings}%
\usepackage[ruled,vlined]{algorithm2e}%
\usepackage[nameinlink,noabbrev]{cleveref}%

\theoremstyle{thmstyleone}%
\theoremstyle{thmstyletwo}%
\newtheorem{remark}{Remark}%

\theoremstyle{thmstylethree}%
\theoremstyle{definition}
\newtheorem{problem}{Problem}[section]

\theoremstyle{plain}
\newtheorem{lemma}{Lemma}[section]

\def\a{{\overline a}}
\def\A{{\overline A}}

\def\w{{\overline w}}
\def\W{{\overline W}}

\def\mA{{\mathbb A}}

\def\mW{{\mathbb W}}

\crefname{problem}{Problem}{Problems}
\Crefname{problem}{Problem}{Problems}
\crefname{lemma}{Lemma}{Lemmas}
\Crefname{lemma}{Lemma}{Lemmas}
\crefname{algocf}{Algorithm}{Algorithms}
\Crefname{algocf}{Algorithm}{Algorithms}

\begin{document}

\title[Industry Aware Firm Level Network Reconstruction]{Industry Aware Firm Level Network Reconstruction}

\author*[1,2]{\fnm{Mitja} \sur{Devetak}}\email{devetak@csh.ac.at}

\author[3,4]{\fnm{Antoine} \sur{Mandel}}\email{antoine.mandel@univ-paris1.fr}

\affil*[1]{\orgname{Complexity Science Hub Vienna}, \orgaddress{\city{Vienna}, \country{Austria}}}

\affil[2]{\orgname{Austrian Supply Chain Intelligence Institute (ASCII)}, \orgaddress{\country{Austria}}}

\affil[3]{\orgname{Centre d'Économie de la Sorbonne, Paris School of Economics, Université Paris 1 Panthéon-Sorbonne}, \orgaddress{\city{Paris}, \country{France}}}

\affil[4]{\orgname{Climate Finance Alpha}, \orgaddress{\city{Paris}, \country{France}}}

\abstract{A number of recent contributions have put forward the topological structure of production networks as a key determinant of macro-economic dynamics. However, firm-to-firm production networks data is generally not available. Against this background, reconstruction method based on firms' size  have been developed. This paper enriches this set of reconstruction methods by integrating input–output sectoral flows in the reconstruction process. We derive analytical expressions for the maximum entropy solutions to the firm network reconstruction problem with sectoral input–output constraints, first for binary networks and then for weight reconstruction. We perform a numerical analysis comparing standard and input–output based reconstruction methods using Hungarian production network data. Our results show that adding input–output constraints substantially reduces deviations from the input–output structure compared with standard methods. Our augmented method provides an almost perfect fit to input–output data, though all methods have difficulties reproducing other structural characteristics.}


\maketitle

\section{Introduction}\label{sec1}

 A number of recent contributions have put forward the topological structure of production networks as a key determinant  of macro-economic dynamics \citep[see e.g.][]{acemoglu2012network,carvalho2014micro,baqaee2019macroeconomic,bigio2020distortions}. A set of empirical contributions has confirmed the quantitative relevance of network effects \citep{barrot2016input,carvalho2021supply,diem2022quantifying} in explaining micro and macro-economic fluctuations. In a context where massive computational power and rich firm-level data are available, this makes a strong case for the development of applied economic models grounded at the firm-level \citep{axtell2018endogenous,dosi2019more,gill2021high,farmer2025quantitative}. However, a key piece of information missing in this perspective is the generic availability of data on firm-to-firm interactions, i.e. firm-to-firm production network data \citep{pichler2023building}. The above contributions  rely on data available for only a handful of countries \citep[see][]{bacilieri2022firm}, which limits the scaling of these models.    

Against this background, a natural approach is to seek to reconstruct firm-level production networks from available firm-level data, i.e. non-relational data such as  sales or revenues. A number of contributions in this perspective have been developed, primarily by adapting techniques from the literature on financial network reconstruction \citep{squartini2018reconstruction,mungo2024reconstructing}. However, this approach neglects a key source of information: data on sectoral flows provided by input–output tables for which increasingly large and comprehensive databases are available \citep[see e.g.][and references therein]{stadler2018exiobase}. 

In this paper, we enrich existing production network reconstruction methods with data  on the value of input–output sectoral flows. We first show how existing methods can be extended to the case where meso-level constraints are present. We derive maximum-entropy solutions for binary reconstruction with sector-pair constraints. We then derive the corresponding weight reconstruction conditional on a sampled topology. For the latter problem, we also provide an extension of the iterative proportional fitting (IPF) method \citep{bacharach1965estimating, squartini2018reconstruction} to the case where meso-scale constraints are present. We then perform a numerical analysis of the performance of standard and input–output based reconstruction methods. 

This analysis shows that integrating input–output constraints is necessary for production network reconstruction, because standard methods otherwise generate large deviations from the input–output structure, which limits their empirical applicability. On the contrary, our augmented method, although probabilistic in nature, provides an almost perfect fit to input–output data. We further show that maximum entropy methods despite theoretical guarantees generate too high a variance for weight reconstruction in the presence of input–output constraints and that the iterative proportional fitting method should be preferred for this use-case. However, all the methods considered have difficulties in reproducing other structural characteristics of the networks such as the shape of the degree distribution. This suggests further extensions of our approach shall be pursued to improve the accuracy of network reconstruction methods.     

The remainder of this paper is structured as follows. We begin with a review of the literature on firm level network reconstruction. We then introduce the reconstruction framework, covering binary models, maximum entropy methods, IPF based approaches, and the CReM class. After this we describe the Hungarian production network data and the harmonisation with the input–output table. The results section evaluates the reconstructed networks against known empirical properties \citep{bacilieri2022firm} and examines how the different methods reproduce firm strengths, industry flows, and structural statistics. We compare Economic Systemic Risk Index (ESRI) values computed on reconstructed networks with values from the original network \citep{diem2022quantifying}. We show that ESRI is misspecified under these reconstructions. Adding industry constraints only improves recovery of aggregate flows, while several micro-level weighted properties remain difficult to match.

\subsection{Literature Review}

Reconstructing networks from partial data is a common problem in many fields \citep[for a review][]{guimera2009missing}. In finance, reconstruction of exposure networks has a long history \citep{squartini2018reconstruction}, including maximum-entropy and IPF-based approaches. More recently, researchers have focused on production networks for economic research \citep{mungo2024reconstructing}, driven by advances in theory \citep{acemoglu2012network} and increased availability of empirically grounded data \citep{pichler2023building}. We contribute by incorporating constraints from aggregate input–output tables. Other approaches \citep{gomez2012inferring, mungo2024reconstructing} require specialized data that is not broadly available.

The idea of using the input–output table in production network reconstruction is not new. Prior work \citep{inoue2019firm, carvalho2021supply} uses industry constraints to assign weights while preserving firm strengths, typically proportional to input–output coefficients. These approaches assume the binary topology is known. We drop that assumption, reconstruct topology and weights, and generate an ensemble rather than a single network. This has significant advantages as it allows for more robust explorations of the phenomena under consideration. Another existing approach is the use of configuration models \citep{hillman2021cab}. In those works they sample random edges and set their weights to the minimum value which would still maintain all three desired constraints, the in-strength of the incoming end, the out strength of the outgoing end and the corresponding entry in the input–output table. In contrast to them our method is able to handle a much larger number of firms, 200.000 against 5000, as it is more computationally efficient. More recent work \citep{ialongo2022reconstructing} also uses input-data, but in that case they have access to by industry inputs from firms, which is not readily available in most cases. Our work is most similar to \citep{fessina2024inferring} with the difference that we exclusively use public data. In all cases, we use the same type of data, firms' in and out strengths, their industry, and the input–output table. Furthermore since we use maximum entropy methods all the assumptions are explicit, which avoids potential biases \citep{rachkov2021potential}.  We will show that despite these beneficial properties the maximum entropy methods under-perform simpler methods we propose.

In this work we focus partially on methods which are based on maximum entropy. We aim to find an ensemble which maximises entropy such that all the known properties of the network are satisfied \citep{squartini2018reconstruction}. We focus on ensembles that enforce constraints in expectation, which keeps the problem numerically tractable. In our setting, the strength constraints and sector-pair flow constraints are imposed in expectation. For constraints subject to measurement error, such as highly aggregated input–output tables, imposing them in expectation can be appropriate \citep{squartini2015unbiased}. Given the ensemble we can sample multiple times and hence get a more robust understanding on the modeled phenomenon. Another benefit of maximum entropy methods is that it makes explicit all assumptions that we use in the network reconstruction. Furthermore, as we will see, it allows to easily add new assumptions to better incorporate domain knowledge. Our approach builds on CReM-B \citep{parisi2020faster}, which has been applied to firm-level reconstruction \citep{bacilieri2023reconstructing}. We extend CReM-B to include input–output table constraints. 
We also propose an alternative weight reconstruction based on iterative proportional fitting (IPF) \citep{bacharach1965estimating}, which is often used to redistribute weights in financial and economical networks \citep{squartini2018reconstruction, lenzen2010uncertainty}. We extend IPF to account for input-output (IO) sectoral constraints and  show that this approach is able to generate more realistic weights than CReM-B for our problem.


\section{Models}

In this section we present the models we will use to reconstruct firm-level network.  Following \citep{parisi2020faster}, the approach is separated in two steps. The first step seeks to reconstruct the adjacency structure of the network: one specifies an ensemble of possible adjacency matrices that satisfies the available information on connectivity in expectation. The second step seeks to reconstruct the weights given the adjacency structure. This separation is required because the joint maximum entropy problem over both topology and weights is computationally intractable in sparse, heterogeneous networks. By treating topology and weights sequentially, one obtains closed form expressions for the weight distributions and tractable likelihood equations for the associated parameters, which form the basis of the CReM family of models. With this in mind we first state the problem of binary reconstruction for firm networks. We present the density corrected gravity model (dcGM) as well as our extension of it which accounts for IO table constraints. Given a sampled binary network and its sampling probabilities, we present two methods to assign weights to edges. First, we introduce IPF and our industry-constrained extension, which we call IPF-industry. Second, we present CReM-B \citep{parisi2020faster} and extend it to incorporate input–output constraints.

\subsection{Notation and General Problem Statement}

We consider as given a set of firms, $\mathcal{F}=\{1,\dots,n\}$ distributed across a set of sectors  $\mathcal{S}.$ Namely, each firm $i\in\mathcal{F}$ is assigned to a sector $S_i\in\mathcal{S}$ and
$\mathcal{F}_S := \{ i\in\mathcal{F} : S_i=S\}$ denotes the set of firms in sector $S.$ We consider that firms are linked by supply relationships represented represented by a  weighted network of monetary flows $\W:=(\w_{ij})_{i,j\in\mathcal{F}}.$ We further denote by $\A:=(\a_{ij})_{i,j\in\mathcal{F}}$ the associated binary adjacency matrix. We assume that $\A$ and $\W$ are not observable but that the following information is available:
\begin{itemize} \label{eq:targets}
\item For each firm  $j$, the out-strength $s_j^{out}$ corresponding to the sum of outgoing flows:
\begin{equation} s_j^{out}=\sum_{j\in\mathcal{F}} \w_{ij}. \label{eq:targets_out}
\end{equation}
In economic terms, this corresponds to total sales to other firms.\\

\item  For each firm  $j$, the in-strength $s_i^{out}$ corresponding to the sum of incoming flows:
\begin{equation} s_j^{in}=\sum_{i\in\mathcal{F}} \w_{ij}  \label{eq:targets_in}\end{equation}
In economic terms this corresponds to total intermediary consumption.\\
\item For each pair of sectors, $(S,B)\in\mathcal{S}^2,$ the total flow from sector $S$ to sector $B$
\begin{equation}    
 s_{S,B}= \sum_{i\in\mathcal{F}_S}\sum_{j\in\mathcal{F}_B} \w_{ij}  \qquad \forall (S,B)\in\mathcal{S}^2. \label{eq:targets_io}
\end{equation}
In economic terms, this corresponds to the entry of the input-output table associated to the corresponding pair of sectors.\\

Furthermore, we shall consider that the average degree (or equivalently the density) of the network is observed or can be set parametrically to a value $k,$ i.e.
\begin{equation} k = \dfrac{1}{n}\sum_{i \in S} \sum_{j \in S} \a_{i,j} \label{eq:mean_degree_target} \end{equation}

\end{itemize}

In general, we shall treat the network reconstruction probabilistically, i.e. we aim to generate a random adjacency matrix $\mathbb{A}$ and an associated random weighted network $\mathbb{W}$ such that, in expectation, the in-strength, the out-strength, the sectoral flows, and the mean degree are consistent with the observations given by Equations \ref{eq:targets_out} to \ref{eq:mean_degree_target}. Formally, our problem can be stated as follows.  

\begin{problem}[Full reconstruction from micro, meso and macro targets]\label{prob:full}
Determine a random adjacency matrix $\mathbb{A}$ and an associated random weighted network $\mathbb{W}$ such that, letting $p_{ij}=\mathbb{P}(\mA_{ij}=1),$ one has:
\begin{enumerate}
  \setcounter{enumi}{-1}

\item The adjacency and weighted network matrix are consistent:
$$\mathbb{P}(\A_{i,j}=0 \Rightarrow \W_{i,j}=0)=1$$  
\item For each $i \in \mathcal{F}$, the expected outgoing flow is: $$\mathbb{E}(\sum_{j \in \mathcal{F}} \mW_{i,j})= s_i^{out}$$
\item For each $j \in \mathcal{F}$,  the expected incoming flow is: $$\mathbb{E}(\sum_{i \in \mathcal{F}} \mW_{i,j})=s_j^{in}$$
\item For each $(S,B)\in\mathcal{S}^2,$ the expected input-output flows is:   
$$ \mathbb{E}(\sum_{i\in\mathcal{F}_S}\sum_{j\in\mathcal{F}_B}\mW_{ij})=s_{S,B} $$ 
\item The expected mean degree is $k$:
$$ \mathbb{E}(\sum_{i=1}^n\sum_{j=1}^n \mA_{ij})=kn,$$
\end{enumerate}
\end{problem}

\begin{remark}
Throughout, constraints are imposed in expectation, i.e.\ the target values define the corresponding
ensemble means. For practical reconstruction one also wants small dispersion around these targets
across realisations. In sparse firm-level production networks, however, each node has only a few
links, so averaging is limited and concentration is weak. The realised degrees and strengths can deviate
substantially from their expectations. This lack of concentration is a key reason why we believe expectation-based
maximum-entropy weight models such as CReM-B can perform poorly in our setting (see Results).
\end{remark}

We address \cref{prob:full} in two stages. First we construct a random adjacency matrix $\mathbb{A}$ such that the expected degree condition $\ref{eq:mean_degree_target}$ holds. We then construct a random weighted network $\mathbb{W}$ such that all the constraints of Problem \ref{prob:full} are satisfied. Though sequential, the two steps are not independent. The adjacency structure $\mathbb{A}$ constructed in the first step must take into account the complete set of constraints in order to ensure that the network reconstruction problem actually has a solution.



\subsection{Binary Network Reconstruction}

In this subsection, we present our reconstruction method for the binary adjacency matrix $\mathbb{A}.$
We focus on methods in which edges are sampled independently, so that the binary adjacency matrix $\mathbb{A}$ is fully characterized by the matrix of marginal probabilities $P=(p_{ij})$ where $p_{ij}=\mathbb{P}(a_{ij}=1).$ 

We first recall the density corrected gravity model (dcGM) from \citep{cimini2015systemic} that builds a solution to the problem of reconstructing the binary adjacency matrix under the expected degree condition $\ref{eq:mean_degree_target}$. More specifically, the dcGM provides a maximum-entropy solution to this problem under a fitness ansatz following which the in and out-strength of the nodes can be used as proxy constraints for the in and out-degree of the nodes \cite[see][for details]{cimini2015systemic}. Namely, the standard dcGM assumes that the probability $p_{ij}$ of a link existing from node $i$ to node $j$ (for $i \not =j$) is given by

\begin{equation}
\label{eq:dcgm}
    p_{ij} = \frac{zs_i^{out}s_j^{in}}{1 + zs_i^{out}s_j^{in}},
\end{equation}
where $z$ is a global parameter which is used to calibrate the expected density. When $i = j$ we set $p_{ij} = 0$ to exclude self-loops. More precisely, given $n$ nodes and a mean degree of $k,$ we choose $z$ such that

\begin{equation}
\label{eq:z_formula}
    \sum_{i = 1}^n\sum_{j = 1}^n p_{ij} = kn.
\end{equation}

\noindent By construction, the dcGM hence satisfies the expected degree condition $\ref{eq:mean_degree_target}.$

\begin{lemma}[dcGM solves the binary reconstruction problem]\label{lem:dcgm}
The random adjacency matrix defined by \cref{eq:dcgm,eq:z_formula} has expected number of links $kn$ and is such that $p_{ii}=0$.
\end{lemma}

As emphasized above, the dcGM method can be understood as a maximum-entropy method with two constraints: the mean degree and a fitness ansatz \citep{PhysRevLett.89.258702, barrat2004architecture} following which degree and strengths are positively correlated. We now state explicitly the maximum-entropy problem. Consider the set of binary adjacency matrices $\mathbb{A}$ with $a_{ii}=0$. We maximise
\begin{equation}
\max_{P}\; -\sum_{A\in\mathbb{A}} P(A)\log P(A)
\end{equation}
subject to $\sum_{A}P(A)=1$, the expected density constraint
\begin{equation}
\sum_{i = 1}^n\sum_{j = 1}^n \mathbb{E}[a_{ij}] = kn,
\end{equation}
and the restriction to an independent-edge ensemble with logit parametrisation
$p_{ij}=\frac{x_{ij}}{1+x_{ij}}$. Under the fitness ansatz
$x_{ij}=z\,s_i^{out}s_j^{in}$ one recovers \cref{eq:dcgm}, while
$x_{ij}=z\,s_{S_i,S_j}s_i^{out}s_j^{in}$ yields \cref{eq:new_dcgm}.
Hence dcGM and dcIAGM are maximum-entropy ensembles under a fitness
restriction and a density constraint.

However, the dcGM does not take into account input-output constraints. For example, two large firms which are in two industries that rarely trade with each other, such as forestry and car manufacturing, will nevertheless be given a high probability of linking. However, the dcGM approach can be extended to account for the sectoral input-output flow constraints.  

Namely, we define the density corrected industry aware gravity model (dcIAGM) by setting the probability $p_{ij}$ when $i \neq j$ to
\begin{equation}
    \label{eq:new_dcgm}
    p_{ij} = \frac{zs_{S_i,S_j}s_i^{out}s_j^{in}}{1 + zs_{S_i, S_j}s_i^{out}s_j^{in}},
\end{equation}
where the $z$ parameter is calibrated according to Equation \eqref{eq:z_formula}, which can be done efficiently with binary search. By construction, the dcIAGM satisfies the expected degree condition $\ref{eq:mean_degree_target}$ and one has:

\begin{lemma}[dcIAGM solves the binary reconstruction problem]\label{lem:dciagm}
The random adjacency matrix defined by \cref{eq:new_dcgm,eq:z_formula} has expected number of links $kn$ and is such that $p_{ii}=0$.
\end{lemma}

In the dcIAGM, the fitness ansatz is extended to account for the strength of inter-sectoral flows on top of the in and out strengths as in the standard dcGM model, as well as the strength of the flow between the two industries. A full derivation is given in Appendix \ref{sec:dciagm}. This extended fitness approach is not unique to firm level production networks and could be applied to any setting in which flows between groups of nodes are known. For example, \citep{ialongo2022reconstructing} modifies the original dcGM model into what they refer to as the stripe corrected dcGM, to account for industry level inputs. This approach was notably used to evaluate the ESRI from reconstruction in \citet{fessina2024inferring}.

\begin{remark}
If partial information about existing nodes in the network is available, one can include it in the modeling approach by setting $p_{ij} = 1$ for the known connections. This is similar to how we set $p_{ii} = 0$ to avoid self loops.
\end{remark}

\section{Weighted Reconstruction}

In this section we assume that a random binary adjacency structure $\mathbb{A}$ has been constructed and focus on the reconstruction of the random weighted network $\mathbb{W}.$ First, we introduce the iterative proportional fitting (IPF) algorithm and extend it to account for inter-sectoral flows. Second, we introduce the state of the art CReM-B approach from \citep{parisi2020faster} and also extend it to account for inter-sectoral flows.

\subsection{Iterative Proportional Fitting}

Under the iterative proportional fitting approach we construct $\mathbb{W}$ by associating to each realization $A$ of $\mathbb{A}$ a weighted network $W$ whose adjacency structure is consistent with $A.$ Specifically, the iterative proportional fitting (IPF) is a matrix scaling procedure that enforces a set of marginal constraints on a non–negative matrix with a fixed support, (see \citealt{bacharach1965estimating,squartini2018reconstruction}). In our setting the support is given by the binary network $A$ and we want the final weighted network to (approximately) satisfy Equations \ref{eq:targets_out} to \ref{eq:mean_degree_target} so as to obtain a (approximate) solution to problem \ref{prob:full}. In this perspective, IPF computes the unique matrix that satisfies the given constraints while maximising Shannon entropy or equivalently minimising the KL divergence from the prior. 

Let $A$ be the binary adjacency matrix, where $a_{ij} \in \{0,1\}$ and $a_{ii} = 0$. The classical IPF algorithm seeks a weighted matrix $W = (w_{ij})$ of the form
\begin{equation}
    w_{ij} = a_{ij} u_i v_j.
\end{equation}
We extend the algorithm to also include industry level factors and in that case the matrix $\mathbb{W}$ is of the form
\begin{equation}
    \label{eq:ipf_factorization}
    w_{ij} = a_{ij} u_i v_j g_{S_i,S_j},
\end{equation}
where $u_i$ and $v_j$ are firm–level scaling factors and $g_{S,B}$ are industry–pair scaling factors. 

As emphasized above, the unknown factors $u_i$, $v_j$ and $g_{S,B}$ are chosen so that out-strengh, in-strength and input-output flow constraints hold:

\begin{align}
    \sum_{j} w_{ij} &= s_i^{out} \quad \text{for all } i,\\
    \sum_{i} w_{ij} &= s_j^{in} \quad \text{for all } j,\\
    \sum_{i \in \mathcal{F}_S}\sum_{j \in \mathcal{F}_B} w_{ij} &= s_{S,B} \quad \text{for all industry pairs } (S,B).
\end{align}


IPF proceeds by alternating multiplicative updates of the scaling factors so that one family of constraints is satisfied at a time. Starting from some initial matrix $W^{(0)} = (w^{(0)}_{ij})$ that respects the support.

\begin{algorithm}[htbp]
\caption{IPF-industry for solving \cref{prob:full}}\label{alg:ipf-ind}
\KwIn{Binary support ${A}$, targets $\{s_i^{out}\}$, $\{s_j^{in}\}$, $\{s_{S,B}\}$, tolerance $\varepsilon$}
\KwOut{Weights ${W}$ supported on ${A}$}

Initialize $w^{(0)}_{ij}\leftarrow a_{ij}$ for all $i,j$ (or any strictly positive prior on the support)\;
\Repeat{all marginal errors or changes $<\varepsilon$}{
  \ForEach{$i$}{
    $w_{ij} \leftarrow \frac{s_i^{out}}{\sum_{\ell} w_{i\ell}}\, w_{ij}$ for all $j$\;
  }
  \ForEach{$j$}{
    $w_{ij} \leftarrow \frac{s_j^{in}}{\sum_{\ell} w_{\ell j}}\, w_{ij}$ for all $i$\;
  }
  \ForEach{sector pair $(S,B)$}{
    $w_{ij} \leftarrow \frac{s_{S,B}}{\sum_{k\in \mathcal{F}_S}\sum_{l\in \mathcal{F}_B} w_{kl}}\, w_{ij}$ for all $(i,j)$ with $(S_i,S_j)=(S,B)$\;
  }
}
\end{algorithm}

\begin{lemma}[IPF-industry solves the weight reconstruction problem]\label{lem:ipf}
Given a feasible support ${A}$, \cref{alg:ipf-ind} outputs a matrix ${W}$ supported on
${A}$ that satisfies the constraints in \cref{eq:targets}. In particular, it provides a
constructive solution to \cref{prob:full}; the resulting ${W}$ admits the multiplicative
form \cref{eq:ipf_factorization}.
\end{lemma}

These three steps are repeated until the changes in the weights or in the marginals are below a given tolerance. We refer to \cref{alg:ipf-ind,lem:ipf} as IPF-industry. The standard IPF variant corresponds to omitting the sector-pair update step and setting $g_{S,B} = 1$ in \cref{eq:ipf_factorization}.

Under standard feasibility conditions the procedure converges to a matrix of the form \eqref{eq:ipf_factorization} that satisfies all constraints and shares the zero pattern of $A$. IPF and IPF-industry are instances of alternating KL-projections \citep{gietl2013accumulation}. This perspective suggests extensions to additional constraints, such as regional input–output tables.

Feasibility requires that the support graph can satisfy the specified marginals \citep{bacharach1965estimating, bacharach1970biproportional}. Unfortunately, since we sample the adjacency matrix we cannot guarantee that these conditions are met. Checking whether a sampled adjacency matrix can satisfy the marginals can be computationally expensive. \cref{alg:ipf-ind} requires (i) each firm with positive out-strength has at least one outgoing edge, (ii) each firm with positive in-strength has at least one incoming edge, and (iii) each sector pair with positive flow has at least one supporting edge between firms in those sectors. Otherwise, some update steps divide by zero, causing the iteration to fail.

We enforce only the necessary conditions above to prevent division by zero. This does not guarantee feasibility for the full set of constraints. We do so by sampling in all cases a single edge proportional to the probabilities given by dcGM and dcIAGM respectively for the firm or sector that requires an edge. This guarantees that the algorithm will return real numbers. Note that this is a necessary but not sufficient condition for the general satisfiability of the IPF algorithms. We find that in practice stopping iterating after the error rate stabilises gives very good results, with most errors concentrated on smaller firms. This procedure slightly increases the realized density relative to the target density because it adds edges when required. Designing a sampling procedure that guarantees feasibility while preserving the target density is left for future work. 

\subsection{CReM-B}

CReM-B assigns weights by maximizing entropy while enforcing constraints in expectation \citep{parisi2020faster}. CReM-B is an alternative approach to \cref{prob:full} that assigns weights to a binary network by maximising Shannon entropy over the distribution of edge weights while enforcing the constraints in \cref{eq:targets} in expectation. In our setting the binary network is represented by a probability matrix $P=(p_{ij})$ obtained from the binary reconstruction (e.g., from dcGM or dcIAGM). Given $P$ and the targets $\{s_i^{out}\}$, $\{s_j^{in}\}$, and $\{s_{S,B}\}$, CReM-B computes a distribution over weight matrices $\mathbb{W}$ supported on $\mathbb{A}$ that satisfies the marginal constraints in expectation.

From entropy maximisation we have that the probability density function of the edge weights conditional of the edge existing is given by an exponential distribution. Conditional on $a_{ij}=1$, weights follow an exponential distribution with rate $\beta_{ij}>0$:
\begin{equation}
\label{eq:weight_distribution}
q(w_{ij}\mid a_{ij}=1)=\beta_{ij}\exp(-\beta_{ij}w_{ij}),\qquad w_{ij}>0.
\end{equation}
Hence $\mathbb{E}[w_{ij}\mid a_{ij}=1]=1/\beta_{ij}$.
As above, we impose $w_{ii}=0$ (no self-loops) by construction.

The algorithm CReM-A, from the same paper, would find $\beta_{ij}$ by maximising the entropy of the weights while enforcing the row and column sums in expectations. By maximising

\begin{equation}
    \max_{W}\; -\sum_{i,j} w_{ij}\ln w_{ij}
\end{equation}

subject to

\begin{equation}
\begin{aligned}
    \sum_j p_{ij} w_{ij} &= s_i^{out},\\
    \sum_i p_{ij} w_{ij} &= s_j^{in},\\
    \sum_{i\in \mathcal{F}_S}\sum_{j\in \mathcal{F}_B} p_{ij} w_{ij} &= s_{S,B},
\end{aligned}
\end{equation}

which yields the system of equations for $\beta_{ij}$ given below.

\begin{equation}
    \label{eq:equation_crem_a}
    \begin{split}
       \quad - s_i^{out} + \sum_{j  = 1}^n \frac{p_{ij}}{\beta_i^{out} + \beta_j^{in} + \beta_{S_i,S_j}} &= 0,\\
        \quad -s_i^{in} + \sum_{j = 1}^n \frac{p_{ji}}{\beta_j^{out} + \beta_i^{in} + \beta_{S_j,S_i}} &= 0,\\
        \quad -s_{S,B} + \sum_{i \in \mathcal{F}_S}\sum_{j \in \mathcal{F}_B}\frac{p_{ij}}{\beta_i^{out} + \beta_j^{in} + \beta_{S,B}} &= 0.
    \end{split}
\end{equation}

where $\beta_{ij}=\beta_i^{out}+\beta_j^{in}+\beta_{S_i,S_j}$ is the rate. The complete derivation is presented in Appendix \ref{sec:derivation:crema}. Unfortunately this is computationally prohibitive as it involves solving $O(n^2)$ simultaneous equations, where $n$ is the number of firms. Hence \citep{parisi2020faster} introduce CReM-B. Equation \eqref{eq:weight_distribution} states that conditional on existence of the edge the weight is distributed exponentially with rate $\beta_{ij}$. 

The second observation from \cite{parisi2020faster} is that:
\begin{equation}
\langle w_{ij}\rangle = p_{ij}\,\mathbb{E}[w_{ij}\mid a_{ij}=1] = \frac{p_{ij}}{\beta_{ij}}.
\end{equation}

If we prescribe a target unconditional mean $\hat w_{ij}$, then we set
\begin{equation}
\beta_{ij}=\frac{p_{ij}}{\hat w_{ij}}.
\end{equation}

Following \citep{parisi2020faster} and building on existing literature \citep{squartini2018reconstruction, mastrandrea2014enhanced}, a natural candidate for $\hat w_{ij}$ is the MaxEnt estimate:

\begin{equation}
    \label{eq:max_ent}
    \hat w_{ij} = \frac{s_i^{out}s_j^{in}}{W^*},
\end{equation}

where $W^*$ is the total flow of the network, $W^* = \sum_{i = 1}^n s_i^{out} = \sum_{i = 1}^n s_i^{in}$. 

\begin{problem}[MaxEnt baseline for expected weights]\label{prob:maxent}
Maximise $-\sum_{i=1}^n\sum_{j=1}^n w_{ij}\ln w_{ij}$ subject to $\sum_{j} w_{ij}=s_i^{out}$ and $\sum_{i} w_{ij}=s_j^{in}$. The solution defines a baseline matrix of expected weights $\widehat{W}=(\hat w_{ij})$, which we later use as target unconditional means in the fitness/CReM-B construction (with $w_{ii}=0$ imposed separately in implementation).
\end{problem}

\begin{lemma}[MaxEnt baseline]\label{lem:maxent}
The unconstrained baseline for \cref{prob:maxent} is given by \cref{eq:max_ent}, where
$W^*=\sum_{i=1}^n s_i^{out}=\sum_{j=1}^n s_j^{in}$. In implementation we then set $w_{ii}=0$ by
construction and do not renormalise.
\end{lemma}

We can then solve for each $\beta_{ij}$ independently and the problem becomes numerically tractable. 

A problem with Equation \eqref{eq:max_ent} is that it assigns non-zero weights to self loops, but before we assigned zero probability to them. Therefore solving for $\beta_{ii}$ would involve division by zero. To solve this we simply set the diagonal weights to zero. This introduces some bias as the row and column sums are no longer as seen in the data, but overall this accounts only for a negligible fraction of the total flows.

Furthermore the MaxEnt algorithm is based on maximum entropy under in and out strengths constraints. Therefore we derive a version which also takes into account the between group of nodes constraints, such are industry level input–output tables, in that case we state the problem as 

\begin{equation}
    \max_{W} \; -\sum_{i=1}^n \sum_{j=1}^n w_{ij}\ln w_{ij}
\end{equation}

subject to

\begin{equation}
\begin{aligned}
    \sum_{j=1}^n w_{ij} &= s_i^{out},\\
    \sum_{i=1}^n w_{ji} &= s_i^{in},\\
    \sum_{i \in \mathcal{F}_S}\sum_{j \in \mathcal{F}_B} w_{ij} &= s_{S,B}.
\end{aligned}
\end{equation}

The solution to \cref{prob:maxent-ind} is given by \cref{lem:maxent-ind}.

\begin{problem}[MaxEnt with sector-pair targets]\label{prob:maxent-ind}
Maximise $-\sum_{i=1}^n\sum_{j=1}^n w_{ij}\ln w_{ij}$ subject to the three sets of constraints in \cref{eq:targets}. The solution defines a baseline matrix of expected weights $\widehat{\mathbb{W}}=(\hat w_{ij})$ (with $w_{ii}=0$ imposed separately in implementation).
\end{problem}

\begin{lemma}[MaxEnt with sector-pair targets]\label{lem:maxent-ind}
The unconstrained baseline for \cref{prob:maxent-ind} is given by \cref{eq:max_ent_ind}. In
implementation we then set $w_{ii}=0$ by construction and do not renormalise.
\end{lemma}

\begin{equation}
    \label{eq:max_ent_ind}
    \hat w_{ij} = \frac{s_{S_i, S_j}s_i^{out}s_j^{in}}{\left(\sum_{k \in \mathcal{F}_{S_i}}s_k^{out}\right)\left(\sum_{l \in \mathcal{F}_{S_j}}s_l^{in}\right)}.
\end{equation}

The derivation of Equation \eqref{eq:max_ent_ind} is available in Appendix \ref{sec:derivation:maxEntWithIndustry}. As expected, if all firms are in one sector we recover Equation \eqref{eq:max_ent}, this is the case as both $s_{S_i, S_j}$ and the two sums in the denominator are equal to $W^*$.

Despite theoretical guarantees this method is not able to effectively reconstruct firm level networks. This is because of the high variance of the network weight reconstruction which is particularly pronounced in the case of sparse networks.

\section{Results}

In this section we assess the performance of the different network reconstruction methods presented against empirical data.  The first part introduces the Hungarian firm data that we use for the reconstruction as well as all the data harmonisation procedures we use between the firm data and the IO table data. Using the four introduced methods we compare how well they perform in reconstructing the known properties of the network, that is the firm in and out strengths as well as the IO table. For each method, we generate 10 networks to estimate sampling variability. Across these samples, the reported statistics vary little (see standard deviations in Tables 1–3). dcIAGM with IPF-industry outperforms the other pipelines in matching sector-pair input–output flows. Next, we compare reconstructed network statistics to the ranges reported in \citep{bacilieri2022firm}. This signals further avenues for research. In the final part of this section we compare the networks on the calculation of the economic systemic risk index (ESRI), a real world application of reconstructed networks.

\subsection{Description of the Data}

We use the 2021 Hungarian firm-level VAT network \citep{borsos2020unfolding} and take firm in-strengths and out-strengths as targets. It is constructed from the same procedure as the 2017 VAT network used by \citep{diem2022quantifying}. When available, each firm has a 4-digit NACE code. We also use ESRI scores computed on the 2021 network. The mean degree in 2021 is 4.54 (source: network summary provided with the data).

The released network is processed: small transactions/links are removed and firms are aggregated at the holding level. This may affect tail and clustering estimates. Following \cite{diem2022quantifying} we classify firms with missing NACE codes as non-essential firms for ESRI computation.

When using the industry level input–output tables we use the OECD 2021 input–output table. We harmonize firm strengths to be consistent with sector totals implied by the table. We treat firms with no NACE code as a separate industry. First, we redistribute sector totals across non-essential industries proportionally to firm strengths within each category. Second, we apply IPF to the input–output table. We modify IPF to keep sector marginals fixed and to keep trade among essential industries fixed. We note that this procedure amounts to assuming that firms with unknown sectors are equally likely to be in any of the non-essential groups in the OECD industry groupings. Input–output accounting does not map one-to-one to supply chain relations \citep{colon2025constructing}. Integrating supply-chain specific IO tables is left for future work.

\subsection{Comparison of known statistics}

In this subsection we compare the reconstruction of the known statistics of the model. We compute the mean out degree of the reconstructed networks as well as the absolute error in various known quantities. To make the absolute errors more understandable we express it as a percentage of the total flow in the network. We also compute flow misalignment which is the percentage difference between the total flow of the real network and the reconstructed networks. Table \ref{tab:degree_io_errors_pct} presents the results.

\begin{table}[htbp]
  \centering
\resizebox{\textwidth}{!}{
\begin{tabular}{lrrrrrrrr}
    \hline
    & \multicolumn{2}{c}{dcGM-CReM}
    & \multicolumn{2}{c}{dcGM-IPF}
    & \multicolumn{2}{c}{dcIAGM-CReM}
    & \multicolumn{2}{c}{dcIAGM-IPF} \\
    Property
    & Mean & Std
    & Mean & Std
    & Mean & Std
    & Mean & Std \\
    \hline
    Mean degree
      & 4.50 & 0.00352
      & 5.50 & 0.00324
      & 5.10 & 0.00470
      & 6.00 & 0.00484 \\
    Total absolute error in flows (in) [\%]
      & 100.0 & 0.189
      & 7.0   & 0.614
      & 110.0 & 0.361
      & 11.0  & 0.325 \\
    Total absolute error in flows (out) [\%]
      & 100.0 & 0.184
      & 7.0   & 0.613
      & 110.0 & 0.304
      & 12.0  & 0.606 \\
    Total absolute error IO table [\%]
      & 110.0 & 0.195
      & 91.0  & 0.573
      & 56.0  & 0.503
      & 4.10  & 0.0782 \\
    Flow misalignment [\%]
      & 0.085 & 0.206
      & 6.50  & 0.538
      & 6.10  & 0.477
      & 3.30  & 0.0742 \\
    \hline
  \end{tabular}
  }
\caption{Mean degree and flow-matching errors across reconstruction pipelines. For each method we report the sample mean and standard deviation over 10 reconstructed networks. ''Total absolute error in flows (in)`` is $\sum_j |\hat s_j^{in}-s_j^{in}|/W^*$ and ''(out)`` is $\sum_i |\hat s_i^{out}-s_i^{out}|/W^*$, expressed in percent, where $W^*=\sum_i s_i^{out}=\sum_j s_j^{in}$. ''Total absolute error IO table`` is $\sum_{S,B} |\widehat s_{S,B}-s_{S,B}|/W^*$ in percent. ''Flow misalignment`` is $|\widehat W^*-W^*|/W^*$ in percent, where $\widehat W^*=\sum_{i,j}\hat w_{ij}$.}
\label{tab:degree_io_errors_pct}
\end{table}

\begin{figure}[ht]
    \centering
    \includegraphics[width=0.8\linewidth]{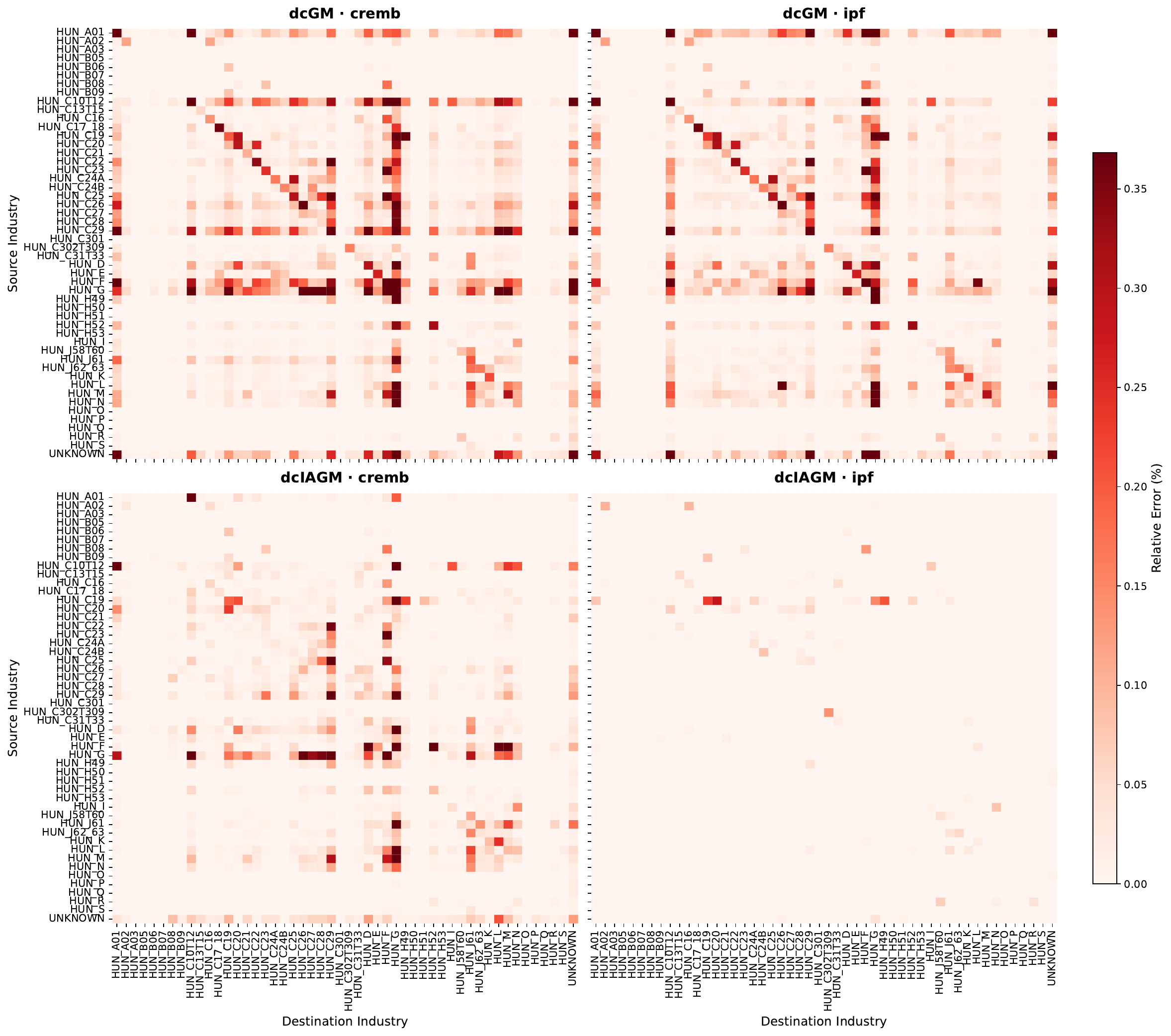}
    \caption{Industry-pair input--output errors for one reconstructed network. Each cell $(S,B)$ shows the relative deviation $(\widehat s_{S,B}-s_{S,B})/s_{S,B}$ between reconstructed and target IO flows from sector $S$ to sector $B$. Cells with $s_{S,B}=0$ are omitted (or set to 0) by construction. Positive values indicate over-allocation of flow from $S$ to $B$; negative values indicate under-allocation.}
\label{fig:error}
\end{figure}

By construction we see that the mean degrees are respected, while we notice a slightly upwards bias in the IPF networks consistent with the ensuring minimal feasibility which always adds edges. We note that without taking into account IO constraints the flows are mischaracterised by $100\%$ for both methods, while in the case firm level flows we show that the IPF based algorithm greatly overperforms the CReM-B algorithms in reconstructing known values of the in and out flows of firms. Similarly for the IO table we get half the error by using our modified version of CReM-B which also accounts for industry flows. This is consistent with us using additional information while reconstructing the weights of the networks. The proposed IPF-Industry algorithm which tries to match the weights exactly greatly over performs the modified version of CReM-B leading to errors which are an order of magnitude smaller. To give visual intuition on the distribution of errors in Figure \ref{fig:error} we present the relative error per industry pair, the sum of which we presented in the table. In reconstructing total flows the dcGM method paired with CReM-B is most consistent while the three other methods showing smaller but non-negligible results.

\subsection{Comparison of network statistics}

Following the results from \citep{bacilieri2022firm} we compute network properties for which there is high or very high evidence in consistency between firm level production networks. For a description of the network properties and how to estimate them we refer to \citep{bacilieri2022firm}. 

For the binary networks we compute the out and in degree tail exponents as well as assortativity, reciprocity and average local clustering. We compute all using the methodology from \citep{bacilieri2022firm} to which we also refer for a more thorough description of the statistics. 

\begin{table}[htbp]
  \centering
\resizebox{\textwidth}{!}{
  \begin{tabular}{llrrrrrrrr}
    \hline
    & & \multicolumn{2}{c}{dcGM-CReM}
    & \multicolumn{2}{c}{dcGM-IPF}
    & \multicolumn{2}{c}{dcIAGM-CReM}
    & \multicolumn{2}{c}{dcIAGM-IPF} \\
    Property & Known
    & Mean & Std
    & Mean & Std
    & Mean & Std
    & Mean & Std \\
    \hline
    Out-degree tail exponent & [1.4, 1.6]
      & 2.00 & 0.046
      & 2.00 & 0.046
      & 2.00 & 0.150
      & 2.00 & 0.013 \\
    In-degree tail exponent & [2, 3]
      & 2.10 & 0.040
      & 2.00 & 0.023
      & 2.10 & 0.120
      & 1.80 & 0.130 \\
    Degree assortativity & [-0.015, -0.13]
      & -0.11 & 0.00048
      & -0.12 & 0.00050
      & -0.12 & 0.00033
      & -0.13 & 0.00023 \\
    Reciprocity [\%] & [4, 5.5]
      & 4.10 & 0.017
      & 3.50 & 0.013
      & 5.70 & 0.028
      & 4.80 & 0.022 \\
    Average local clustering [\%] & [20, 28]
      & 25.0 & 0.067
      & 24.0 & 0.070
      & 30.0 & 0.140
      & 21.0 & 0.067 \\
    \hline
  \end{tabular}
  }
\caption{Binary-network statistics of reconstructed ensembles compared to empirical ranges reported by \citet{bacilieri2022firm}. For each method we report mean and standard deviation over 10 samples. Tail exponents are estimated on in- and out-degree distributions; assortativity is the Pearson correlation between degrees at the ends of directed edges; reciprocity is the fraction of edges that are bidirectional; average local clustering is computed on the directed graph using the same definition as \citet{bacilieri2022firm}.}
\label{tab:degree_rec_clust}
\end{table}

In Table \ref{tab:degree_rec_clust} we present the results as well as the known values from the literature. All methods perform roughly equally on all statistics without a superior winner.

Similarly for a weighted network we can compute the tail exponent of the tail as well as the various OLS slopes between degrees and strengths as well as the influence vector tail exponents. We find that for the tails exponent of the weights all methods perform comparably and outside of the expected range. For the OLS slope we find that CReM-B based models were notably better at reconstructing the known correlations from the data. We also note the huge discrepancy in the influence vector tail exponent which we attribute to the data cleaning procedure performed by the Bank of Hungary.

\begin{table}[htbp]
  \centering
   \resizebox{\textwidth}{!}{
  \begin{tabular}{llrrrrrrrr}
    \hline
    & & \multicolumn{2}{c}{dcGM-CReM}
    & \multicolumn{2}{c}{dcGM-IPF}
    & \multicolumn{2}{c}{dcIAGM-CReM}
    & \multicolumn{2}{c}{dcIAGM-IPF} \\
    Property & Known
    & Mean & Std
    & Mean & Std
    & Mean & Std
    & Mean & Std \\
    \hline
    Weights tail exponent & [1.1, 1.2]
      & 2.60 & 0.019
      & 2.50 & 0.035
      & 2.40 & 0.250
      & 2.40 & 0.014 \\
    Out-strength $\sim$ out-degree (OLS slope) & [0.9, 1.05]
      & 1.20 & 0.0021
      & 1.30 & 0.0073
      & 1.10 & 0.0035
      & 1.20 & 0.0500 \\
    Out-degree $\sim$ out-strength (OLS slope) & [0.31, 0.36]
      & 0.64 & 0.0020
      & 0.040 & 0.00073
      & 0.47 & 0.0016
      & 0.025 & 0.0028 \\
    In-strength $\sim$ in-degree (OLS slope) & [1.35, 1.54]
      & 1.20 & 0.0012
      & 1.80 & 0.0120
      & 1.10 & 0.0029
      & 2.10 & 0.0950 \\
    In-degree $\sim$ in-strength (OLS slope) & [0.40, 0.45]
      & 0.62 & 0.00051
      & 0.015 & 0.00023
      & 0.46 & 0.0014
      & 0.017 & 0.00045 \\
    Influence vector tail exponent & [1.05, 1.4]
      & 680  & 270
      & 8600 & 1900
      & 660  & 150
      & 6000 & 3000 \\
    \hline
  \end{tabular}
  }
\caption{Weighted-network statistics of reconstructed ensembles compared to empirical ranges from \citet{bacilieri2022firm}. For each method we report mean and standard deviation over 10 samples. ``Weights tail exponent'' is the estimated tail exponent of the edge-weight distribution. The four OLS slopes quantify strength--degree scaling (log--log regressions as in \citet{bacilieri2022firm}). ``Influence vector tail exponent'' is computed from the influence (Leontief-inverse-based) measure using the same estimator as \citet{bacilieri2022firm}.}
\label{tab:weights_strength_influence}
\end{table}

Overall we conclude that the binary network reconstruction is comparable across models. While our proposed methodology shines in reconstruction macro properties it clearly lacks in the reconstruction of micro level weighted properties. In particular it fails to accurately capture the correlation relationships between strengths and degrees. Making the correlation mostly flat.

\subsection{Comparison to ESRI results}

ESRI is a systemic risk indicator which assigns to each firm a number between $0$ to $1$. This value can be interpreted as the percentage of the economy which is dependent, both as customer and supplier, on that firm. Hence, a firm with ESRI score $0.1$ can be interpreted as : a disruption to this firm will reduce economic output by $10\%$. The ESRI computation is performed as described in \citep{diem2022quantifying} for brevity we only provide a brief self-contained description of the method in the Appendix \ref{sec:esri}. Further results are presented in Appendix \ref{sec:esri_additional}.

We find that the reconstructed ESRI scores are higher in synthetic networks with an average ESRI of $5\cdot10^{-4}$ against an empirical average of $3\cdot10^{-5}$. Since ESRI is used to identify small but systemically important firms in the economy this mismatch does not invalidate the reconstruction. Therefore in Table \ref{tab:topx_seedwise_mean} we compare the top firms identified by ESRI and review how many of the top firms are found similar as it was done in \citep{fessina2024inferring}. We find that all methods are very poor at recovering the most systemically important firms. We note that all methods perform similarly only recovering a third of the $100$ most systemically important Hungarian firms and having much lower percentages over the top. In fact only 5 networks ever recovered one of the $10$ most important firms.

\begin{table}[ht]
\centering
\begin{tabular}{lcccc}
\hline
Method & Top 10 & Top 20 & Top 50 & Top 100 \\
\hline
dcGM-IPF      & 0.0 & 4.6 & 18.0 & 36.8 \\
dcGM-CReM      & 0.3 & 7.0 & 16.8 & 35.8 \\
dcIAGM-IPF     & 0.1 & 4.1 & 18.8 & 37.4 \\
dcIAGM-CReM   & 0.0 & 6.0 & 17.8 & 37.2 \\
\hline
\end{tabular}
\caption{Average number of recovered systemically important firms from the Hungarian data using the computed ESRI over all realizations of the network per method.}
\label{tab:topx_seedwise_mean}
\end{table}

Comparing to prior work \citet{fessina2024inferring} we recover less of the original ESRI, in their work they report $7$ out of the top $10$ firms were recovered. This is due to the fact that we only use publicly available input-output tables, which are not from aggregate firm level flows, see \citet{colon2025constructing}, as well as down-scale the firm industries to 4 categories only after reconstruction since we need to match the input-output table we are given. In Table \ref{tab:topx_max_over_seeds} we also compare the same metric where we take for each firm the maximum realized ESRI we note that the results do not change much. We think this result shows the limitation of imperfect information on ESRI computations.

\begin{table}[ht]
\centering
\begin{tabular}{lcccc}
\hline
Method & Top 10 & Top 20 & Top 50 & Top 100 \\
\hline
dcGM-IPF       & 0 & 3 & 17 & 40 \\
dcGM-CReM     & 0 & 6 & 17 & 36 \\
dcIAGM-IPF     & 0 & 2 & 17 & 36 \\
dcIAGM-CReM & 0 & 6 & 17 & 34 \\
\hline
\end{tabular}
\caption{Number of recovered systemically important firms from the Hungarian data using the maximum computed ESRI over all realizations of the network per method.}
\label{tab:topx_max_over_seeds}
\end{table}

\section{Conclusion}

We develop industry-aware firm-level reconstruction methods that incorporate input–output constraints. We derive maximum-entropy solutions for binary and weighted reconstruction, and we propose practical pipelines combining dcGM (topology) with IPF-based weight assignment. Using Hungarian firm data harmonized with an OECD input–output table, we show that standard approaches can deviate strongly from sector-pair flows. Adding sector-pair constraints matches input–output flows closely. However, micro-level statistics and ESRI rankings remain difficult to reproduce. At the same time, the results highlight remaining gaps in reproducing micro-level network statistics and ESRI rankings, pointing to the need for further extensions that better capture firm-level heterogeneity beyond the constraints enforced here.

\backmatter

\bmhead{Author contributions}
 MD and AM designed the research. MD developed the model, analyzed the results, and wrote the paper. AM provided feedback on the model, results, and paper.

\bmhead{Acknowledgements}
MD and AM thank András Borsos for providing the data. MD also thanks Peter Klimek, Christian Diem, and Kjartan van Driel for helpful comments on the text. 

\bmhead{Funding}
AM acknowledges funding from the EU Horizon Europe program through the project ST4TE (Strategies for just and equitable transitions in Europe) grant number 101132559.

\bmhead{Competing interests}
The authors declare no competing interests.

\bmhead{Data and Code Availability}

The firm level data is sensitive and hence not available publicly. The code for the network reconstruction is freely available at \url{https://github.com/Devetak/network_reconstruction}. The code for computing the Economy Systemic Risk Index of an economy is freely available at \url{https://github.com/Devetak/ESRI.jl}


\begin{appendices}

\section{Derivation of Density Corrected Models}
\label{sec:dciagm}

We provide a self-contained maximum entropy derivation of the binary model used for dcIAGM. Let $\mathbb{A}=(a_{ij})_{i,j\in\mathcal{F}}$ be a directed adjacency matrix
with $a_{ij}\in\{0,1\}$, and impose $a_{ii}=0$. We seek an ensemble over $\mathbb{A}$
that maximises Shannon entropy subject to the available information used in
\cref{prob:full,eq:mean_degree_target}. Concretely, we maximise
\begin{equation}
\label{eq:binary_entropy}
\max_{P}\;\; -\sum_{A\in \mathbb{A}} P(A)\log P(A),
\end{equation}
subject to $\sum_{A\in\mathbb{A}}P(A)=1$ and constraints imposed in expectation. In addition to the
density target \cref{eq:mean_degree_target}, we encode the ``fitness'' ansatz used in the main text
by restricting to independent-edge ensembles with link probabilities of the form
\begin{equation}
\label{eq:binary_ansatz}
p_{ij}=\mathbb{P}(a_{ij}=1)=\frac{x_{ij}}{1+x_{ij}},\qquad i\neq j,\qquad p_{ii}=0,
\end{equation}
for some nonnegative intensities $x_{ij}$. The role of the maximum entropy principle is then to
select the exponential-family distribution over $A$ consistent with \cref{eq:binary_ansatz} and the
density constraint.

Define the Hamiltonian
\begin{equation}
\label{eq:binary_hamiltonian}
H(A)=\sum_{i\neq j}\theta_{ij}\,a_{ij},
\end{equation}
and consider the maximum-entropy distribution of the form
\begin{equation}
\label{eq:binary_gibbs}
P(A)=\frac{e^{-H(A)}}{Z},\qquad Z=\sum_{A\in\mathbb{A}} e^{-H(A)}.
\end{equation}
Because $H(A)$ is additive over dyads, \cref{eq:binary_gibbs} factorises over $(i,j)$ and yields
independent Bernoulli variables with probabilities
\begin{equation}
\label{eq:binary_logit}
p_{ij}=\frac{e^{-\theta_{ij}}}{1+e^{-\theta_{ij}}},\qquad i\neq j,\qquad p_{ii}=0.
\end{equation}
Identifying $x_{ij}:=e^{-\theta_{ij}}$ shows that \cref{eq:binary_logit} is exactly
\cref{eq:binary_ansatz}.

As in \citet{cimini2015systemic}, we interpret the parameters $x_{ij}$ through a fitness ansatz in which observable node characteristics act as proxies for degree heterogeneity. In directed networks, strengths are empirically correlated with degrees, so one may set $x_{ij}\propto s_i^{out}s_j^{in}$, which yields the dcGM specification. In our setting we extend this fitness ansatz by including a sector-pair factor that reflects the intensity of inter-industry flows observed in the input-output table. This multiplicative correction rescales the odds of links between firms in sectors that trade heavily while preserving the same maximum-entropy independent-edge structure.

We now specify $x_{ij}$ using the same data and notation as in the main text. The dcIAGM choice is
\begin{equation}
\label{eq:binary_xij_dciagm}
x_{ij}=z\,s_{S_i,S_j}\,s_i^{out}\,s_j^{in},\qquad i\neq j,
\end{equation}
where $z>0$ is a global parameter controlling density. Substituting
\cref{eq:binary_xij_dciagm} into \cref{eq:binary_ansatz} (equivalently, into \cref{eq:binary_logit})
yields
\begin{equation}
p_{ij}
=
\frac{z\,s_{S_i,S_j}\,s_i^{out}\,s_j^{in}}{1+z\,s_{S_i,S_j}\,s_i^{out}\,s_j^{in}},
\qquad i\neq j,
\qquad p_{ii}=0,
\end{equation}
which is \cref{eq:new_dcgm}. The density target \cref{eq:mean_degree_target} fixes $z$ by requiring
\begin{equation}
\sum_{i=1}^n\sum_{j=1}^n p_{ij}=kn,\qquad p_{ii}=0,
\end{equation}
which is \cref{eq:z_formula}.

Finally, to recover dcGM, assume all firms belong to the same sector. Then $s_{S_i,S_j}$ is constant
across all pairs $(i,j)$; absorbing this constant into $z$ reduces the above expression to
\begin{equation}
p_{ij}=\frac{z\,s_i^{out}\,s_j^{in}}{1+z\,s_i^{out}\,s_j^{in}},
\qquad i\neq j,
\qquad p_{ii}=0,
\end{equation}
which is \cref{eq:dcgm}.

\section{Derivation of CReM-A Method for Input-Output Table Constrains}
\label{sec:derivation:crema}

This is a direct derivation from \citep{parisi2020faster} with the addition that we now consider also input–output table constraints. We fix the in and out strengths as well as the input table constraints as

\begin{equation}
    \begin{split}
    \quad s_i^{out} = \sum_{j = 1}^n w_{ij},\\
    \quad s_i^{in} = \sum_{j = 1}^n w_{ji},\\
    \quad s_{S, B} = \sum_{i \in \mathcal{F}_S}\sum_{j \in \mathcal{F}_B} w_{ij}.
    \end{split}
\end{equation}

with these constraints the Hamiltonian from Equation (III.11) becomes

\begin{equation}
    H(W) = \sum_{i = 1}^n\left[\beta_i^{out}s_i^{out}(W) + \beta_i^{in}s_i^{in}(W)\right] + \sum_{S}\sum_{B} \beta_{S,B} s_{S, B}(W).
\end{equation}

Note that the Hamiltonian is defined directly for the whole network $W$. Furthermore we introduced the functional notation that $s_i^{out}(W) = \sum_{j = 1}^n w_{ij}$ and similar for $s_i^{in}(W)$ and $s_{S, B}(W)$, this is to avoid writing excessive sums. Furthermore the partition function from Equation (III.12) becomes

\begin{equation}
\begin{split}
    Z_{P, \beta}
                            &= \prod_{i=1}^{n}\prod_{j = 1}^n \left[\int_{0}^{\infty} e^{-\left(\beta_i^{out} + \beta_j^{in} + \beta_{S_i, S_j}\right)w_{ij}}\,dw_{ij}\right]^{p_{ij}}\\
                            &= \prod_{i=1}^{n}\prod_{j = 1}^n\left(\frac{1}{\beta_i^{out} + \beta_j^{in} + \beta_{S_i, S_j}}\right)^{p_{ij}}.
\end{split}
\end{equation}

Where $p_{ij}$ is the probability of a link existing from $i$ to $j$ as in the main text. By combining this new Hamiltionian and partition function with Equation (III.6) from the original paper we get a changed version of Equation (III.14) with $q_{ij}(w = 0|a_{ij} = 0) = 1$ and

\begin{equation}
    q_{ij}(w|a_{ij} = 1) = \begin{cases}
        (\beta_i^{out} + \beta_j^{in} + \beta_{S_i, S_j})e^{-(\beta_i^{out} + \beta_j^{in} + \beta_{S_i, S_j})w}\quad &w>0,\\
        0 \quad &w \leq 0.
    \end{cases}
\end{equation}

Where $q_{ij}$ is the distribution of the weights $w_{ij}$ conditional on the realisation $a_{ij}$ of the Bernoulli variable that is defined by $p_{ij}$. This is Equation \eqref{eq:weight_distribution} in the main text. Now we need to derive the equations for the $\beta$s. In this case the new generalised likelihood, Equations (III.7) and (III.15), becomes

\begin{equation}
    \begin{split}
           &G(\beta) = - H_\beta(\langle W\rangle) - \sum_{A \in \mathbb{A}}P(A)\log Z_{A, \beta},\\
  &= - \sum_{i = 1}^{n}\left(s_i^{out}\beta_i^{out} + s_i^{in}\beta_i^{in}\right) - \sum_{S}\sum_B\left(s_{S,B}\beta_{S,B}\right) + \sum_{i = 1}^n\sum_{j = 1}^n p_{ij}\log \left(\beta_i^{out} + \beta_j^{in} + \beta_{S_i, S_j}\right),
    \end{split}
\end{equation}

where $A \in \mathbb{A}$ indicates all the possible adjacency matrices that are possible and $P(A)$ is the probability of their realisation. The last step is to maximise the generalised likelihood. We do so by finding a stationary point. We therefore recover Equation \eqref{eq:equation_crem_a} from the main text.

\begin{equation}
    \tag{\ref{eq:equation_crem_a}}
    \begin{split}
       \quad &- s_i^{out} + \sum_{j  = 1}^n \frac{p_{ij}}{\beta_i^{out} + \beta_j^{in} + \beta_{S_i,S_j}} = 0,\\
        \quad &-s_i^{in} + \sum_{j = 1}^n \frac{p_{ji}}{\beta_j^{out} + \beta_i^{in} + \beta_{S_j,S_i}} = 0,\\
        \quad &-s_{S,B} + \sum_{i \in \mathcal{F}_S}\sum_{j \in \mathcal{F}_B}\frac{p_{ij}}{\beta_i^{out} + \beta_j^{in} + \beta_{S,B}} = 0.
    \end{split}
\end{equation}

We can then use a numerical solver to solve the system. We found that in practice convergence is slow and the fact that one needs to store the entirety of the probability matrix as well as re-computing the factors in the sum at each iteration makes the cost of computations high. Furthermore, the structure of the equations makes them a poor candidate for hardware acceleration.

\section{Derivation of Equation \eqref{eq:max_ent_ind}}

\label{sec:derivation:maxEntWithIndustry}

As in the case of MaxEnt we try to maximise the entropy of the weights defined as

\begin{equation}
    S(W) = -\sum_{i = 1}^n\sum_{j = 1}^n w_{ij} \ln w_{ij},
\end{equation}

such that both the in and out constraints for all firm $i$ are satisfied

\begin{equation}
    \sum_{j = 1}^n w_{ij} = s_i^{out} \quad \sum_{j = 1}^n w_{ji} = s_i^{in},
\end{equation}

for all $i$. This would give the classical MaxEnt. We also include the industry constraints, in this case

\begin{equation}
    \sum_{i \in \mathcal{F}_S}\sum_{j \in \mathcal{F}_B} w_{ij} = s_{S,B},
\end{equation}

for all industry pairs $S,B$.

Introducing Lagrange multipliers for the three sets of constraints and taking derivatives of the corresponding Lagrangian with respect to each $w_{ij}$ gives the first order condition
$$
-(1 + \ln w_{ij}) + \lambda_i + \mu_j + \nu_{S_i,S_j} = 0.
$$

Hence we have that

\begin{equation}
    w_{ij} = e^{\lambda_i + \mu_j + \nu_{S_i, S_j}},
\end{equation}

where $S_i$ and $S_j$ represent the sectors in which firm $i$ and $j$ are respectively. In this case we note that the total flow, denoted by $W^* = \sum_{i = 1}^n s_i^{in} = \sum_{i = 1}^n s_i^{out}$, is equal to

\begin{equation}
    W^* = \sum_{i = 1}^n \sum_{j = 1}^n w_{ij} = \sum_{S} \sum_{i \in \mathcal{F}_S} \sum_{B} \sum_{j \in \mathcal{F}_B} e^{\lambda_i + \mu_j + \nu_{S_i, S_j}}.
\end{equation}

Hence

\begin{equation}
    W^* = \sum_{S}  \sum_{i \in \mathcal{F}_S} e^{\lambda_i} \sum_{B} e^{ \nu_{S,B}} \sum_{j \in \mathcal{F}_B} e^{ \mu_j}.
\end{equation}

So that

\begin{equation}
    \frac{W^*}{\sum_i e^{\lambda_i}} = \sum_{B} e^{\nu_{S,B}} \sum_{j \in \mathcal{F}_B} e^{ \mu_j}.
\end{equation}

We also note that similarly we can get

\begin{equation}
    s_i^{out} = \sum_{j = 1}^n w_{ij} = \sum_{j = 1}^n e^{\lambda_i + \mu_j + \nu_{S_i, S_j}} = e^{\lambda_i} \sum_{B} e^{\nu_{S_i, B}} \sum_{j \in \mathcal{F}_B} e^{\mu_j}.
\end{equation}

Hence we have that combining the two equations yields

\begin{equation}
    s_i^{out} = e^{\lambda_i} \frac{W^*}{\sum_{i' = 1}^n e^{\lambda_i'}}.
\end{equation}

From which we conclude that

\begin{equation}
    \label{eq:lambda_MaxEnt}
    e^{\lambda_i} = s_i^{out}\,\frac{\sum_{i' = 1}^n e^{\lambda_{i'}}}{W^*}.
\end{equation}

By a similar procedure we get that

\begin{equation}
    \label{eq:mu_MaxEnt}
    e^{\mu_i} = s_i^{in}\,\frac{\sum_{i' = 1}^n e^{\mu_{i'}}}{W^*}.
\end{equation}

Furthermore we notice that

\begin{equation}
    s_{S,B} = \sum_{i \in \mathcal{F}_S} \sum_{j \in \mathcal{F}_B} e^{\lambda_i + \mu_j + \nu_{S, B}} = e^{\nu_{S,B}} \sum_{i \in \mathcal{F}_S}\sum_{j \in \mathcal{F}_B}e^{\lambda_i + \mu_j} = e^{\nu_{S,B}} \sum_{i \in \mathcal{F}_S}e^{\lambda_i}\sum_{j \in \mathcal{F}_B}e^{ \mu_j} .
\end{equation}

Hence using Equation \eqref{eq:lambda_MaxEnt} and \eqref{eq:mu_MaxEnt} we get

\begin{equation}
    s_{S,B} = \sum_{i \in \mathcal{F}_S} \sum_{j \in \mathcal{F}_B} e^{\lambda_i + \mu_j + \nu_{S, B}} = e^{\nu_{S,B}} \sum_{i \in \mathcal{F}_S}\sum_{j \in \mathcal{F}_B}e^{\lambda_i + \mu_j} = e^{\nu_{S,B}} \sum_{i \in \mathcal{F}_S}e^{\lambda_i}\sum_{j \in \mathcal{F}_B}e^{ \mu_j} .
\end{equation}

This implies that

\begin{equation}
     e^{\nu_{S,B}} = \frac{s_{S,B}}{\sum_{i \in \mathcal{F}_S}e^{\lambda_i}\sum_{j \in \mathcal{F}_B}e^{ \mu_j}}.
\end{equation}

Which using Equations \eqref{eq:lambda_MaxEnt} and \eqref{eq:mu_MaxEnt} we can rewrite as

\begin{equation}
     e^{\nu_{S,B}} = \frac{s_{S,B}}{\left(\sum_{i \in \mathcal{F}_S}s_i^{out} W^* \sum_{i' = 1}^n e^{\lambda_i'}\right)\left(\sum_{j \in \mathcal{F}_B}s_j^{in} W^* \sum_{
     j' = 1}^n e^{\mu_j'}\right)}.
\end{equation}

Finally we can put it all together to get

\begin{equation}
    w_{ij} = \left(e^{\nu_{S_i, S_j}}\right)\left(e^{\lambda_i}\right)\left(e^{\mu_j}\right)
\end{equation}

transforms to

\begin{equation}
    \left(\frac{s_{S_i,S_j}}{\left(\sum_{i' \in \mathcal{F}_{S_i}}s_{i'}^{out} W^* \sum_{i'' = 1}^n e^{\lambda_i''}\right)\left(\sum_{j' \in \mathcal{F}_{S_j}}s_{j'}^{in} W^* \sum_{
     j'' = 1}^n e^{\mu_j''}\right)}\right)\left(s_i^{out} W^* \sum_{i' = 1}^n e^{\lambda_i'}\right)\left( s_j^{in} W^* \sum_{i' = 1}^n e^{\mu_i'}\right).
\end{equation}

Which simplifies to Equation \eqref{eq:max_ent_ind}.

\begin{equation}
    \hat w_{ij} = \frac{s_{S_i, S_j}s_i^{out}s_j^{in}}{\left(\sum_{k \in \mathcal{F}_{S_i}}s_k^{out}\right)\left(\sum_{l \in \mathcal{F}_{S_j}}s_l^{in}\right)}. \tag{\ref{eq:max_ent_ind}}
\end{equation}

\section{Definition of ESRI}
\label{sec:esri}
We assume that firms produce a single output type. Let $\Pi_{i, S}$ indicate the amount of output from industry $S$ that is delivered to firm $i$, let $\alpha_{i,S}$, $\alpha_i$ and $\beta_i$ be parameters related to the production function which are calibrated. Furthermore, let $\mathcal{S}_i^{es}$ indicate the set of essential industries to the production of firm $i$ and let $\mathcal{S}_i^{ne}$ indicate the set of non-essential industries to the production of firm $i$. We assume that firms produce good $x_i$ using the following production function

\begin{equation}
    \label{eq:general_leontief_production_function}
    x_i = \min \left( \min_{S \in \mathcal{S}_i^{es}} \left( \frac{1}{\alpha_{i, S}}\Pi_{i,S}\right), \beta_i + \frac{1}{\alpha_i}\sum_{S \in \mathcal{S}_i^{ne}} \Pi_{i, S}\right),
\end{equation}

which is referred to as generalised Leontief production function. This assumes that firms produce with essential inputs, which are Leontief, and non-essential inputs, which are linear. Missing an essential input will stop production, while missing an inessential input will reduce the input proportionally to its use in the production function. The production function is calibrated based on the production network. Firms that are part of the physical producing industries, NACE class between 1 and 43, have all physical producing industries as essential and non-physical producing industries, NACE class from 45 to 99, as non-essential. Firm that are not part have only non-essential inputs. We use the production network disaggregated to $4$ NACE digits to compute the equilibrium. Note that 

\begin{equation}
    \Pi_{i, S} = \sum_{j \in \mathcal{F}_S} w_{ji}.
\end{equation}

We use 4 digit NACE codes in the production function. The economic systemic risk index of a firm $i$ is the share of the total output that would be lost if firm $i$ were to cease to exist. Notice that there are two ways in which shocks propagate. Downstream, by firms that supply from firm $i$ not getting the necessary inputs and hence being unable to meet demand and upstream by firms that supply to firm $i$ losing a customer and hence needing to produce less. We don't directly model rewiring, the changing of suppliers, but we do model substitution of inputs within a certain industry. To measure how replaceable a firm is we take the current market share defined as

\begin{equation}
    \sigma_j(t) = \min\left(\frac{s_j^{out}(0)}{\sum_{i \in \mathcal{F}_{S_j}} s_i^{out}(t)}, 1\right),
\end{equation}

where $(0)$ indicates the initial state and $(t)$ indicates the current step in the propagation of the shock. That is a firm that has a high market share will be hard to replace, while a firm that has a small market share will be easier to replace. Note that $s_j^{out}(t)$ can only decrease with the number of steps $t$. We propagate the shock until a new equilibrium is reached. Despite substitution in practice the downstream component of the shock is much more significant, especially as essential industries have a large part of their inputs in the Leontief part of the production function. The full derivation of the dynamics together with many more details can be found in \citep{diem2022quantifying}.

\section{Additional ESRI results}
\label{sec:esri_additional}

\begin{figure}[ht]
    \centering
    \includegraphics[width=0.8\linewidth]{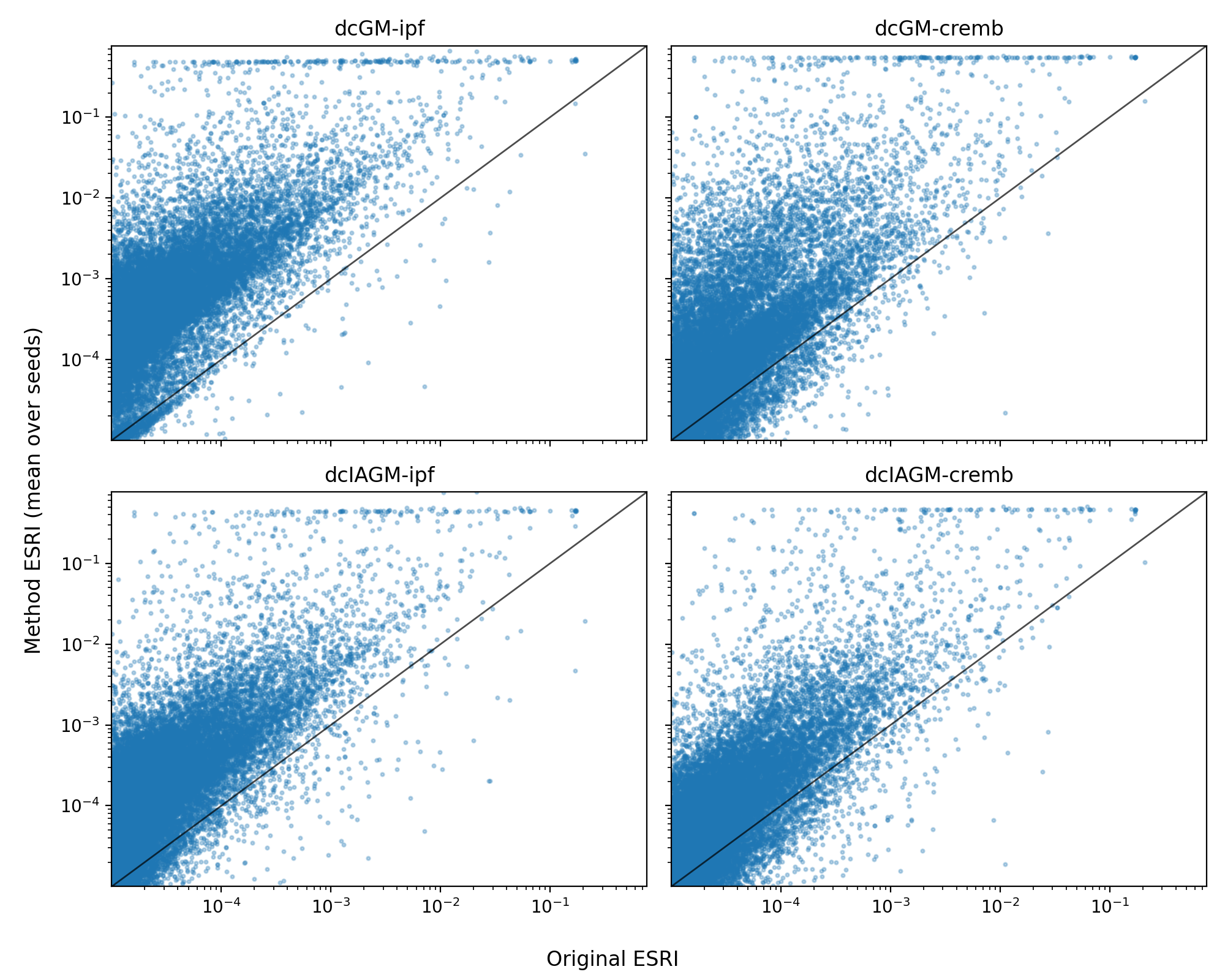}
    \caption{Comparison of Economic Systemic Risk Index (ESRI) values between the empirical network and reconstructed networks across four methods. Each panel shows a log-log scatter plot of empirical ESRI (x-axis) versus the mean over $10$ realizations of method ESRI (y-axis) for one reconstruction method. Points below $10^{-5}$ on either axis are omitted. The diagonal line ($y=x$) indicates equality between empirical and reconstructed ESRI.}
    \label{fig:esri_mean}
\end{figure}

\begin{figure}[ht]
    \centering
    \includegraphics[width=0.8\linewidth]{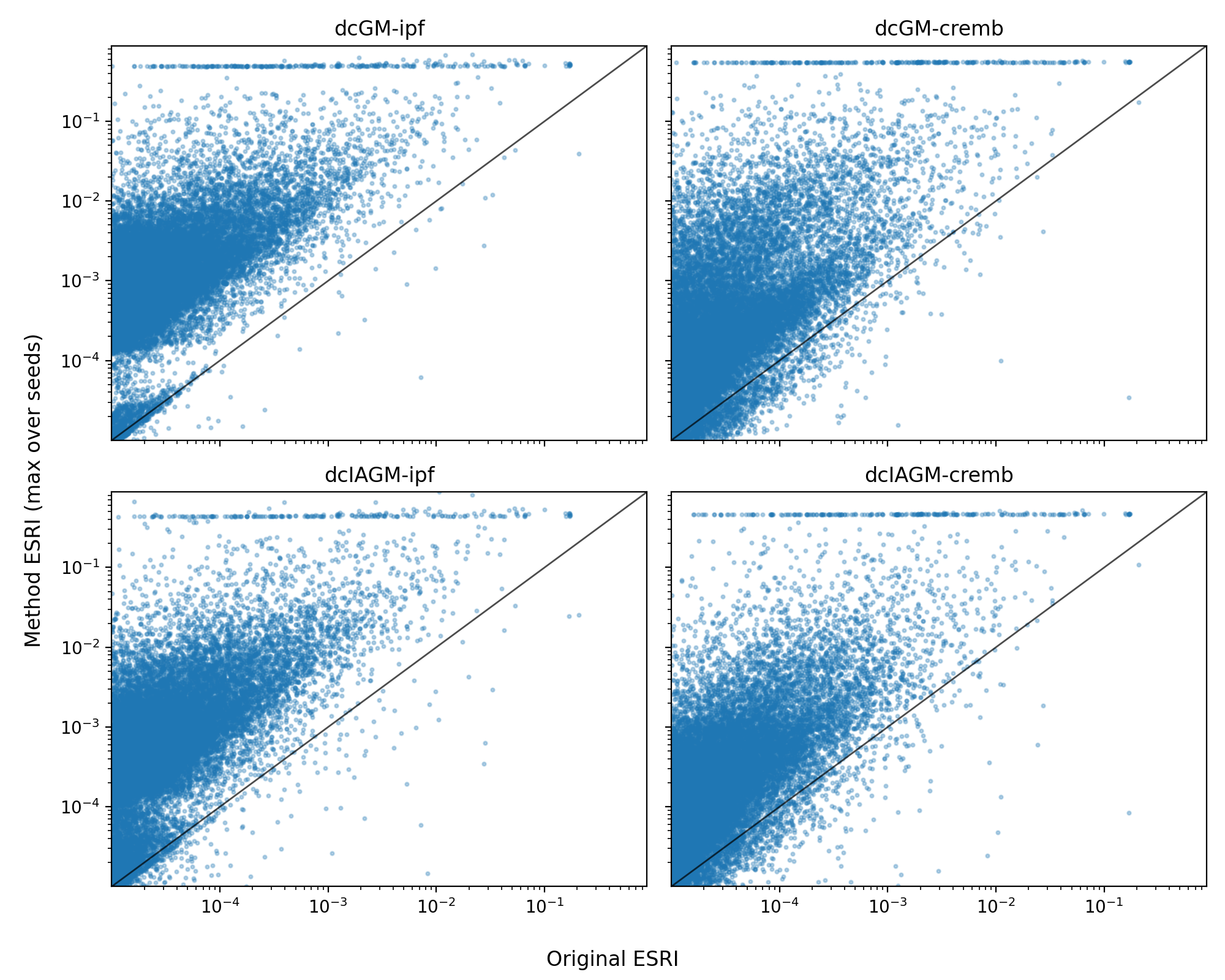}
    \caption{Comparison of Economic Systemic Risk Index (ESRI) values between the empirical network and reconstructed networks across four methods. Each panel shows a log-log scatter plot of empirical ESRI (x-axis) versus the maximum over $10$ realizations of method ESRI (y-axis) for one reconstruction method. Points below $10^{-5}$ on either axis are omitted. The diagonal line ($y=x$) indicates equality between empirical and reconstructed ESRI.}
    \label{fig:esri_max}
\end{figure}

\begin{figure}[ht]
    \centering
    \includegraphics[width=0.8\linewidth]{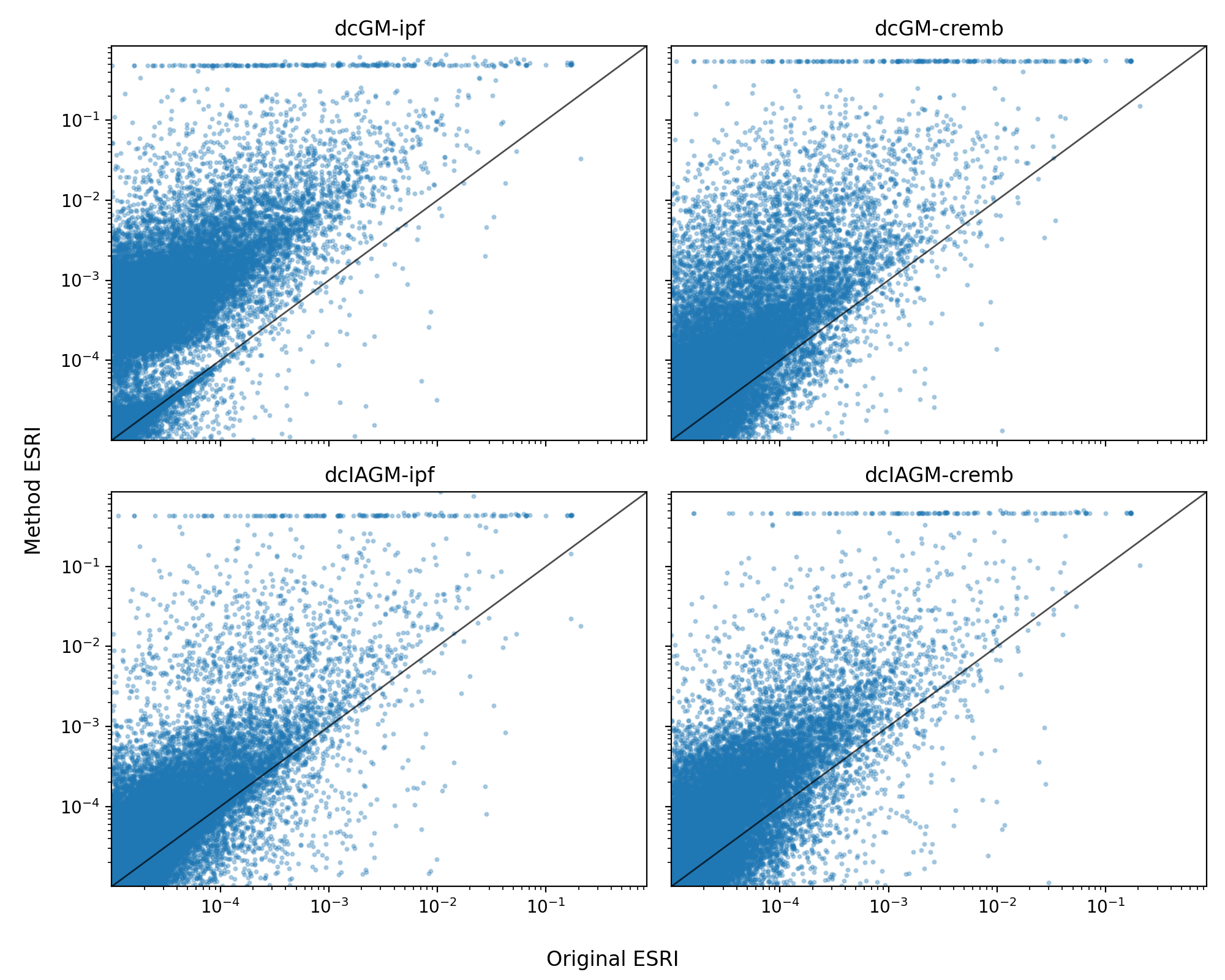}
    \caption{Comparison of Economic Systemic Risk Index (ESRI) values between the empirical network and reconstructed networks across four methods. Each panel shows a log-log scatter plot of empirical ESRI (x-axis) versus method ESRI (y-axis) for a single realization for one reconstruction method. Points below $10^{-5}$ on either axis are omitted. The diagonal line ($y=x$) indicates equality between empirical and reconstructed ESRI.}
    \label{fig:esri_single}
\end{figure}

This Appendix reports ESRI--ESRI comparisons between the empirical network and reconstructed networks. Figure \ref{fig:esri_mean} compares empirical ESRI to the mean ESRI over $10$ realizations for each method. Figure \ref{fig:esri_max} compares empirical ESRI to the maximum ESRI over $10$ realizations for each method. Figure \ref{fig:esri_single} reports the same comparison for a single realization. In all panels, the diagonal line $y=x$ is the equality benchmark. Points below $10^{-5}$ on either axis are omitted. The empirical ESRI and simulated ESRI are also compared in levels. 

\end{appendices}
\end{document}